\documentclass[11pt]{article}
\usepackage[english]{babel}
\usepackage{graphicx,subfigure}
\usepackage{amssymb,amsthm,amsmath,amsfonts}
\newtheorem{theorem}{Theorem}[section]

\newtheorem{definition}[theorem]{Definition}
\newtheorem{example}[theorem]{Example}
\allowdisplaybreaks

\usepackage{float}
\usepackage[colorlinks=true,linkcolor=blue,citecolor=red, urlcolor=green]{hyperref} 
\usepackage{cite}

\begin{document}

\title{Consistent approximate Q-conditional symmetries of PDEs: application to a hyperbolic reaction-diffusion-convection equation}

\author{M.~Gorgone, F.~Oliveri\\
\ \\
{\footnotesize Department of Mathematical and Computer Sciences,}\\
{\footnotesize Physical Sciences and Earth Sciences, University of Messina}\\
{\footnotesize Viale F. Stagno d'Alcontres 31, 98166 Messina, Italy}\\
{\footnotesize mgorgone@unime.it; foliveri@unime.it}
}

\date{Published in \emph{Zeitschrift f\"ur Angewandte Mathematik und Physik (ZAMP) \textbf{72}, 119 (2021).}}

\maketitle

\begin{abstract}
Within the theoretical framework of a recently introduced approach to approximate Lie symmetries of differential equations 
containing small terms, which is consistent with the principles of perturbative analysis, we define accordingly 
approximate Q-conditional symmetries of partial differential equations. The approach is illustrated by considering the hyperbolic version of a 
reaction-diffusion-convection equation. By looking for its first order approximate Q-conditional symmetries, we are able to explicitly determine a large set of non-trivial  approximate solutions.
\end{abstract}

Reaction-Diffusion-Convection Equations; Hyperbolic equations.


\section{Introduction}
A general algorithmic approach to the investigation of differential equations is provided by groups of continuous transformations 
introduced in the nineteenth century by Sophus Lie \cite{Lie1,Lie2}; they are characterized by their infinitesimal generators, and establish a diffeomorphism on the space of independent and dependent variables, mapping solutions of the equations into other solutions (see the monographies \cite{Ovsiannikov,Ibragimov,Olver1,Olver2,Ibragimov:CRC,Baumann,BlumanAnco,Meleshko2005,BlumanCheviakovAnco,Bordag}, or the review paper \cite{Oliveri:Symmetry}). Any transformation of the independent and dependent variables reflects in a transformation of the derivatives by requiring that the contact conditions are preserved. Looking for the Lie group of point transformations leaving a differential equation invariant consists in solving a \emph{linear} system of determining equations for the components of its infinitesimal generators. Lie's theory allows for  the development of systematic procedures leading to the integration by quadrature (or at least to lowering the order) of ordinary differential equations, to the determination of invariant solutions of initial and boundary value problems, to the derivation of conserved quantities, or to the construction of relations between different differential equations that turn out to be equivalent \cite{BlumanAnco,BlumanCheviakovAnco,Oliveri:Symmetry,DonatoOliveri1994,DonatoOliveri1995,DonatoOliveri1996,GorgoneOliveri1,GorgoneOliveri2,Oliveri:quasilinear}.

In the last decades, many extensions and generalizations of Lie's approach have been introduced, and a wide set 
of applied problems has been successfully faced. Here, we focus on conditional symmetries whose origin dates back to 1969 with a seminal paper by Bluman and Cole 
\cite{Bluman-Cole-1969} on the \emph{nonclassical method}, where the continuous symmetries mapping the manifold characterized by the linear heat equation and the invariant surface condition (together with its differential consequences) into itself has been considered. 

This idea has been extended in \cite{Fuschich,Fuschich-Tsyfra}, where the general concept of conditional invariance has been stated. The nonclassical method of Bluman and Cole, nowadays called the method of Q-conditional symmetries, consists in replacing the constraints for the invariance of the given differential equations by the weak conditions of the invariance of the combined system made by the original equations, the invariant surface conditions and their differential consequences. As a result, fewer determining equations, which are \emph{nonlinear}, arise, and in principle a larger set of symmetries can be recovered. Even if the nonlinear determining equations rarely can be solved in their generality, it is possible to find some particular solutions allowing us to determine non-trivial solutions of the differential equations occurring in concrete applications. The available literature on this subject is
quite large  (see, for instance, \cite{Arrigo,Clarkson-Kruskal,Clarkson-Mansfield,Levi-Winternitz,Nucci1993,NucciClarkson1992,Olver-Rosenau-1,Olver-Rosenau-2,Saccomandi});  many recent applications of Q-conditional symmetries to reaction-diffusion equations can be found in \cite{Cherniha-JMAA2007,Cherniha-JPA2010,ChernihaDavydovych-MCM2011,ChernihaDavydovych-CNSNS2012,ChernihaPliukhin-JPA2008}. 

Furthermore, in applied problems it is not uncommon to deal with differential equations containing terms of different orders of magnitude; the occurrence of small terms has the effect of destroying many useful symmetries,  and this restricts the set of invariant solutions that can be found.
This limitation still remains when looking for conditional symmetries.  To overcome this inconvenient, some \emph{approximate symmetry theories} have been proposed, and the notion of  \emph{approximate invariance} properly defined. 

The first approach was introduced by Baikov, Gazizov and Ibragimov \cite{BGI-1989} (see also \cite{IbragimovKovalev}), 
who in 1988 proposed to expand in a perturbation series the Lie generator in order to have
an approximate generator; this approach has been applied to many physical situations 
\cite{BaikovKordyukova2003,DolapciPakdemirli2004,GazizovIbragimovLukashchuk2010,GazizovIbragimov2014,Gan_Qu_ND2010,%
IbragimovUnalJogreus2004,Kara_ND2008,Kovalev_ND2000,Pakdemirli2004,Wiltshire1996,Wiltshire2006}. 
The main problem of this approach is that the expanded generator is not consistent with the principles of perturbation analysis 
\cite{Nayfeh} because the dependent variables are not expanded; as a consequence, in some examples the approximate invariant solutions that are found with this method are not the most general ones. A second approach has been proposed in 1989 by Fushchich and Shtelen \cite{FS-1989}: the dependent variables are expanded in a  series as done in usual perturbation analysis,
terms are then separated at each order of approximation, whence a system of equations to be solved in a hierarchy is obtained.  This resulting system is assumed to be coupled, and the approximate symmetries of the original equation are defined as the \emph{exact symmetries} of the equations obtained from perturbations.
This approach, besides requiring a lot of algebra, is not completely satisfactory since it does not provide a full mixing of perturbative analysis and Lie's method; applications of this method to various equations can be found, for instance, in the papers \cite{Wiltshire2006,Diatta,Euler1,Euler2,Euler3}. 

In a recent paper \cite{DSGO-lieapprox}, an approximate symmetry theory which is consistent with the basic tenets of perturbation analysis, and such that the relevant properties of exact Lie symmetries of differential equations are inherited, has been proposed. More precisely, the dependent variables are expanded in power series of the small parameter as done in classical perturbation analysis; then, instead of considering the approximate symmetries as the exact symmetries of the approximate system (as done in Fushchich-Shtelen method), the consequent expansion of the Lie generator is constructed, and the approximate invariance  with respect to an approximate Lie generator is introduced, as in Baikov-Gazizov-Ibragimov method. An application of the method to the equations of a creeping flow of a second grade fluid can be found in \cite{Gorgone-IJNLM2018}.

For differential equations containing small terms, approximate Q-conditional symmetries can be considered as well (see, for instance, \cite{Mahomed2000,Shih2005}, dealing with applications within the theoretical framework proposed by Baikov, Gazizov and Ibragimov).

The plan of the paper is as follows. In Section~\ref{sec2}, after fixing the notation, we review the approach of approximate symmetries introduced in \cite{DSGO-lieapprox} and accordingly define approximate Q-conditional symmetries of differential equations. In Section~\ref{sec3}, we illustrate the method by computing the first order approximate Q-conditional symmetries of the hyperbolic version of a reaction-diffusion-convection equation, and find various classes of approximate solutions. Some preliminary results about Q-conditional symmetries of the equation herein considered have been given in \cite{GorgoneOliveri-EJDE}, where very special forms of the infinitesimal generators have been chosen. Finally, Section~\ref{sec4} contains  our conclusions.

\section{Theoretical setting}
\label{sec2}
Let us consider an $r$th order differential equation containing small terms, say
\begin{equation}
\label{approxDE}
\Delta\left(\mathbf{x},\mathbf{u},\mathbf{u}^{(r)};\varepsilon\right)=0,
\end{equation}
where $\mathbf{x}\equiv(x_1,\ldots,x_n)\in \mathcal{X}\subseteq \mathbb{R}^n$ are the independent variables, $\mathbf{u}\equiv(u_1,\ldots,u_m)\in \mathcal{U}\subseteq \mathbb{R}^m$ the dependent variables, $\mathbf{u}^{(r)}$ denotes the set of derivatives up to the order $r$ of the $\mathbf{u}$'s with respect to the $\mathbf{x}$'s, and  $\varepsilon\ll 1$ is a constant positive parameter. To keep the paper self-contained, let us briefly review the theory of approximate symmetries according to the approach introduced in \cite{DSGO-lieapprox}.

In perturbation theory \cite{Nayfeh}, differential equations like \eqref{approxDE} involving small terms  
are often studied by looking for solutions in the form
\begin{equation}
\label{expansion_u}
\mathbf{u}(\mathbf{x};\varepsilon)=\sum_{k=0}^p\varepsilon^k \mathbf{u}_{(k)}(\mathbf{x})+O(\varepsilon^{p+1}),
\end{equation}
whose insertion in \eqref{approxDE} provides
\begin{equation}
\Delta\equiv \sum_{k=0}^p\varepsilon^k\widetilde{\Delta}_{(k)}\left(\mathbf{x},\mathbf{u}_{(0)},\mathbf{u}^{(r)}_{(0)},
\ldots,\mathbf{u}_{(k)},\mathbf{u}^{(r)}_{(k)}\right)=O(\varepsilon^{p+1}).
\end{equation}
Now, let us consider a Lie generator
\begin{equation}
\Xi=\sum_{i=1}^n\xi_i(\mathbf{x},\mathbf{u};\varepsilon)\frac{\partial}{\partial x_i}
+\sum_{\alpha=1}^m\eta_\alpha(\mathbf{x},\mathbf{u};\varepsilon)\frac{\partial}{\partial u_\alpha},
\end{equation}
where we assume that the infinitesimals  depend on the small parameter $\varepsilon$.

By using the expansion~\eqref{expansion_u} of the dependent variables only, we have for the infinitesimals
\begin{equation}
\xi_i\approx\sum_{k=0}^p\varepsilon^k \widetilde{\xi}_{(k)i}, \qquad \eta_\alpha\approx\sum_{k=0}^p\varepsilon^k\widetilde{\eta}_{(k)\alpha},
\end{equation}
where $f \approx g$ means $f-g=O(\varepsilon^{p+1})$,
\begin{equation}
\begin{aligned}
&\widetilde{\xi}_{(0)i}=\xi_{(0)i}=\xi_i(\mathbf{x},\mathbf{u}_{(0)};0),\qquad
\widetilde{\eta}_{(0)\alpha}=\eta_{(0)\alpha}=\eta_\alpha(\mathbf{x},\mathbf{u}_{(0)};0),\\
&\widetilde{\xi}_{(k+1)i}=\frac{1}{k+1}\mathcal{R}[\widetilde{\xi}_{(k)i}],\qquad \widetilde{\eta}_{(k+1)\alpha}=\frac{1}{k+1}\mathcal{R}[\widetilde{\eta}_{(k)\alpha}],
\end{aligned}
\end{equation}
$\mathcal{R}$ being a \emph{linear} recursion operator satisfying the \emph{product rule} of derivatives and such that
\begin{equation}
\label{R_operator}
\begin{aligned}
&\mathcal{R}\left[\frac{\partial^{|\tau|}{f}_{(k)}(\mathbf{x},\mathbf{u}_{(0)})}{\partial u_{(0)1}^{\tau_1}\dots\partial u_{(0)m}^{\tau_m}}\right]=\frac{\partial^{|\tau|}{f}_{(k+1)}(\mathbf{x},\mathbf{u}_{(0)})}{\partial u_{(0)1}^{\tau_1}\dots\partial u_{(0)m}^{\tau_m}}
+\sum_{i=1}^m\frac{\partial}{\partial u_{(0)i}}\left(\frac{\partial^{|\tau|} {f}_{(k)}(\mathbf{x},\mathbf{u}_{(0)})}{\partial u_{(0)1}^{\tau_1}\dots\partial u_{(0)m}^{\tau_m}}\right)u_{(1)i},\\
&\mathcal{R}[u_{(k)j}]=(k+1)u_{(k+1)j},
\end{aligned}
\end{equation}
where $k\ge 0$,  $j=1,\ldots,m$, $|\tau|=\tau_1+\cdots+\tau_m$.

Thence, we have an approximate Lie generator
\begin{equation}
\Xi\approx \sum_{k=0}^p\varepsilon^k\widetilde{\Xi}_{(k)},
\end{equation}
where
\begin{equation}
\widetilde{\Xi}_{(k)}=\sum_{i=1}^n\widetilde{\xi}_{(k)i}(\mathbf{x},\mathbf{u}_{(0)},\ldots,\mathbf{u}_{(k)})
\frac{\partial}{\partial x_i}
+\sum_{\alpha=1}^m\widetilde{\eta}_{(k)\alpha}(\mathbf{x},\mathbf{u}_{(0)},\ldots,\mathbf{u}_{(k)})\frac{\partial}{\partial u_\alpha}.
\end{equation}

Since we have to deal with differential equations, we need to prolong the Lie generator
to account for the transformation of derivatives. This is done as in classical Lie group analysis of differential equations,
\emph{i.e.}, the derivatives are transformed in such a way the contact conditions are preserved. 
Of course, in the expression of prolongations, we need to take into account the expansions of $\xi_i$, $\eta_\alpha$ and $u_\alpha$,  and drop the $O(\varepsilon^{p+1})$ terms.

For instance, taking $p=1$, we have the first order approximate Lie generator
\begin{equation}\label{gen_approx}
\begin{aligned}
\Xi &\approx \sum_{i=1}^n\left(\xi_{(0)i}+\varepsilon\left(
 \xi_{(1)i}+\sum_{\beta=1}^m\frac{\partial \xi_{(0)i}}{\partial u_{(0)\beta}}u_{(1)\beta}\right)\right)\frac{\partial}{\partial x_i}\\
&+\sum_{\alpha=1}^m\left(\eta_{(0)\alpha}+\varepsilon\left(
 \eta_{(1)\alpha}+\sum_{\beta=1}^m\frac{\partial \eta_{(0)\alpha}}{\partial u_{(0)\beta}}u_{(1)\beta}\right)\right)\frac{\partial}{\partial u_\alpha},
\end{aligned}
\end{equation}
where $\xi_{(0)i}$, $\xi_{(1)i}$, $\eta_{(0)\alpha}$ and $\eta_{(1)\alpha}$ depend on $(\mathbf{x},\mathbf{u}_{(0)})$.
The first order prolongation is
\begin{equation}
\Xi^{(1)}\approx\Xi + \sum_{\alpha=1}^m\sum_{i=1}^n \eta_{\alpha,i}\frac{\partial}{\partial \frac{\partial u_\alpha}{\partial x_i}},
\end{equation}
where
\begin{equation}
\begin{aligned}
\eta_{\alpha,i} &= \frac{D}{D x_i}\left(\eta_{(0)\alpha}+\varepsilon\left(
 \eta_{(1)\alpha}+\sum_{\beta=1}^m\frac{\partial \eta_{(0)\alpha}}{\partial u_{(0)\beta}}u_{(1)\beta}\right)\right)\\
 &-\sum_{j=1}^n \frac{D}{D x_i}\left(\xi_{(0)j}+\varepsilon\left(
 \xi_{(1)j}+\sum_{\beta=1}^m\frac{\partial \xi_{(0)j}}{\partial u_{(0)\beta}}u_{(1)\beta}\right)\right)
 \left(\frac{\partial u_{(0)\alpha}}{\partial x_j}+\varepsilon \frac{\partial u_{(1)\alpha}}{\partial x_j}\right),
\end{aligned}
\end{equation}
along with the \emph{approximate} Lie derivative
\begin{equation}
\frac{D}{Dx_i}=\frac{\partial}{\partial x_i}+\sum_{k=0}^p\left(\sum_{\alpha=1}^m \left(\frac{\partial u_{(k)\alpha}}{\partial x_i}\frac{\partial}{\partial u_{(k)\alpha}}+ \sum_{j=1}^n\frac{\partial^2 u_{(k)\alpha}}{\partial x_i\partial x_j}\frac{\partial}{\partial\left(\frac{\partial u_{(k)\alpha}}{\partial x_j}\right)}+\cdots \right)\right).
\end{equation}
Similar reasonings lead to higher order prolongations.

The approximate (at the order $p$) invariance condition of the differential equation \eqref{approxDE} reads:
\begin{equation}\label{inv_cond_approx}
\left.\Xi^{(r)}(\Delta)\right|_{\Delta=O(\varepsilon^{p+1})}= O(\varepsilon^{p+1}).
\end{equation}
In the resulting condition we have to insert the expansion of $\mathbf{u}$ in order to obtain the determining 
equations at the various orders in $\varepsilon$. The integration of the determining equations provides the 
infinitesimal generators of the admitted approximate symmetries, and approximate invariant solutions may be 
determined by solving the approximate invariant surface conditions
\begin{equation}
\mathbf{Q}\equiv \sum_{k=0}^p\varepsilon^k\left(\sum_{i=1}^n\widetilde{\xi}_{(k)i}(\mathbf{x},\mathbf{u}_{(0)},\ldots,\mathbf{u}_{(k)})
\frac{\partial \mathbf{u}}{\partial x_i}-
\widetilde{\boldsymbol \eta}_{(k)}(\mathbf{x},\mathbf{u}_{(0)},\ldots,\mathbf{u}_{(k)})\right)=O(\varepsilon^{p+1}),
\end{equation}
where $\mathbf{u}$ is expanded as in \eqref{expansion_u}.

The Lie generator $\widetilde{\Xi}_{(0)}$ is always a symmetry of the unperturbed equations ($\varepsilon=0$); the  \emph{correction}
term $\displaystyle\sum_{k=1}^p\varepsilon^k\widetilde{\Xi}_{(k)}$ gives
the deformation of the symmetry due to the terms involving $\varepsilon$. 
Nevertheless, not all symmetries of the unperturbed equations are admitted as the zeroth terms of 
approximate symmetries; the symmetries
of the unperturbed equations that are the zeroth terms of approximate symmetries are called \emph{stable 
symmetries} \cite{BGI-1989}.  

Let us show the following example, where all the steps of the computation of first order approximate Lie symmetries have been detailed.

\begin{example}
\label{ex:simple}
Let us consider the linear telegraph equation in (1 + 1) dimensions
\begin{equation}\label{telegraph}
\Delta\equiv\varepsilon \frac{\partial^2 u}{\partial t^2}+\frac{\partial u}{\partial t}-\frac{\partial^2 u}{\partial x^2}=0,
\end{equation}
and apply the algorithm described above in order to determine the approximate Lie symmetries corresponding to the approximate generator
\begin{equation}
\begin{aligned}
\Xi &\approx 
\left(\tau_{(0)}(t,x,u_{(0)})+\varepsilon\left(
 \tau_{(1)}(t,x,u_{(0)})+\frac{\partial \tau_{(0)}(t,x,u_{(0)})}{\partial u_{(0)}}u_{(1)}\right)\right)\frac{\partial}{\partial t}\\
 &+\left(\xi_{(0)}(t,x,u_{(0)})+\varepsilon\left(
 \xi_{(1)}(t,x,u_{(0)})+\frac{\partial \xi_{(0)}(t,x,u_{(0)})}{\partial u_{(0)}}u_{(1)}\right)\right)\frac{\partial}{\partial x}\\
&+\left(\eta_{(0)}(t,x,u_{(0)})+\varepsilon\left(
 \eta_{(1)}(t,x,u_{(0)})+\frac{\partial \eta_{(0)}(t,x,u_{(0)})}{\partial u_{(0)}}u_{(1)}\right)\right)\frac{\partial}{\partial u}.
\end{aligned}
\end{equation}
By inserting the expansion of the dependent variable $u(t,x;\varepsilon)$ at first order in $\varepsilon$, that is,
\begin{equation}
u(t,x;\varepsilon)=u_{(0)}(t,x)+\varepsilon u_{(1)}(t,x)+O(\varepsilon^2),
\end{equation}
and using Equation \eqref{telegraph} in \eqref{inv_cond_approx}, \emph{i.e.},
\begin{equation}
\begin{aligned}
&\frac{\partial^2 u_{(0)}}{\partial x^2}=\frac{\partial u_{(0)}}{\partial t},\\
&\frac{\partial^2 u_{(1)}}{\partial x^2}=\frac{\partial^2 u_{(0)}}{\partial t^2}+\frac{\partial u_{(0)}}{\partial t},
\end{aligned}
\end{equation}
we obtain the approximate invariance condition:
{\footnotesize
\begin{align*}
&2\frac{\partial \tau_{(0)}}{\partial u_{(0)}}\frac{\partial^2 u_{(0)}}{\partial t \partial x}\frac{\partial u_{(0)}}{\partial x}+2\frac{\partial \tau_{(0)}}{\partial x}\frac{\partial^2 u_{(0)}}{\partial t \partial x}+\frac{\partial^2 \xi_{(0)}}{\partial u_{(0)}^2}\left(\frac{\partial u_{(0)}}{ \partial x}\right)^3+\frac{\partial^2 \tau_{(0)}}{\partial u_{(0)}^2}\frac{\partial u_{(0)}}{ \partial t}\left(\frac{\partial u_{(0)}}{ \partial x}\right)^2\\
&+\left(2\frac{\partial^2 \xi_{(0)}}{\partial x \partial u_{(0)}}-\frac{\partial^2 \eta_{(0)}}{\partial u_{(0)}^2}\right)\left(\frac{\partial u_{(0)}}{ \partial x}\right)^2+2\left(\frac{\partial^2 \tau_{(0)}}{\partial x \partial u_{(0)}}+\frac{\partial \xi_{(0)}}{\partial u_{(0)}}\right)\frac{\partial u_{(0)}}{ \partial t}\frac{\partial u_{(0)}}{ \partial x}\\
&+\left(\frac{\partial^2 \tau_{(0)}}{\partial x^2 }-\frac{\partial \tau_{(0)}}{\partial t }+2\frac{\partial \xi_{(0)}}{\partial x}\right)\frac{\partial u_{(0)}}{ \partial t}+\left(\frac{\partial^2 \xi_{(0)}}{\partial x^2 }-2\frac{\partial^2 \eta_{(0)}}{\partial x \partial u_{(0)}}-\frac{\partial \xi_{(0)}}{\partial t}\right)\frac{\partial u_{(0)}}{ \partial x}-\frac{\partial^2 \eta_{(0)}}{\partial x^2}+\frac{\partial \eta_{(0)}}{\partial t}\\
&+\varepsilon\left(-2\frac{\partial \tau_{(0)}}{\partial u_{(0)}}\frac{\partial^2 u_{(0)}}{ \partial {t}^2}\frac{\partial u_{(0)}}{ \partial t}+2\frac{\partial \xi_{(0)}}{\partial u_{(0)}}\frac{\partial^2 u_{(0)}}{ \partial {t}^2}\frac{\partial u_{(0)}}{ \partial x}-2\left(\frac{\partial \tau_{(0)}}{\partial t}-\frac{\partial \xi_{(0)}}{\partial x}\right)\frac{\partial^2 u_{(0)}}{ \partial {t}^2} \right.\\
&-2\frac{\partial \xi_{(0)}}{\partial u_{(0)}}\frac{\partial^2 u_{(0)}}{ \partial {t} \partial x}\frac{\partial u_{(0)}}{ \partial t}+2\left(\frac{\partial^2 \tau_{(0)}}{\partial u_{(0)}^2}u_{(1)}+\frac{\partial \tau_{(1)}}{\partial u_{(0)}}\right)\frac{\partial^2 u_{(0)}}{ \partial {t} \partial x}\frac{\partial u_{(0)}}{ \partial x}+2\frac{\partial \tau_{(0)}}{\partial u_{(0)}}\frac{\partial^2 u_{(0)}}{ \partial {t} \partial x}\frac{\partial u_{(1)}}{ \partial x}\\
&+2\left(\frac{\partial^2 \tau_{(0)}}{\partial x \partial u_{(0)}}u_{(1)}-\frac{\partial \xi_{(0)}}{\partial t} +\frac{\partial \tau_{(1)}}{\partial x}  \right)\frac{\partial^2 u_{(0)}}{ \partial {t} \partial x}+2\frac{\partial \tau_{(0)}}{\partial u_{(0)}}\frac{\partial u_{(1)}}{ \partial {t}\partial x}\frac{\partial u_{(0)}}{ \partial x}+2\frac{\partial \tau_{(0)}}{\partial x}\frac{\partial u_{(1)}}{ \partial {t}\partial x}\\
&-\frac{\partial^2 \tau_{(0)}}{\partial u_{(0)}^2}\left(\frac{\partial u_{(0)}}{ \partial t}\right)^3-\frac{\partial^2 \xi_{(0)}}{\partial u_{(0)}^2}\left(\frac{\partial u_{(0)}}{ \partial t}\right)^2\frac{\partial u_{(0)}}{ \partial x}+\left(\frac{\partial^3 \tau_{(0)}}{\partial u_{(0)}^3}u_{(1)}+ \frac{\partial^2 \tau_{(1)}}{\partial u_{(0)}^2} \right) \left(\frac{\partial u_{(0)}}{ \partial x}\right)^2\frac{\partial u_{(0)}}{ \partial t}\\
&+\left(\frac{\partial^3 \xi_{(0)}}{\partial u_{(0)}^3}u_{(1)}+ \frac{\partial^2 \xi_{(1)}}{\partial u_{(0)}^2} \right)\left(\frac{\partial u_{(0)}}{ \partial x}\right)^3+2\frac{\partial^2 \tau_{(0)}}{\partial u_{(0)}^2}\frac{\partial u_{(0)}}{ \partial t}\frac{\partial u_{(0)}}{ \partial x}\frac{\partial u_{(1)}}{ \partial x}+\frac{\partial^2 \tau_{(0)}}{\partial u_{(0)}^2}\left(\frac{\partial u_{(0)}}{ \partial x}\right)^2\frac{\partial u_{(1)}}{ \partial t}\\
&+3\frac{\partial^2 \xi_{(0)}}{\partial u_{(0)}^2}\left(\frac{\partial u_{(0)}}{ \partial x}\right)^2\frac{\partial u_{(1)}}{ \partial x}-\left(2\frac{\partial^2 \tau_{(0)}}{\partial t \partial u_{(0)}}-\frac{\partial^2 \eta_{(0)}}{\partial u_{(0)}^2}\right)\left(\frac{\partial u_{(0)}}{ \partial t}\right)^2\\
&+2\left(\left(\frac{\partial^3 \tau_{(0)}}{\partial x\partial u_{(0)}^2}+\frac{\partial^2 \xi_{(0)}}{\partial u_{(0)}^2}\right)u_{(1)}-\frac{\partial^2 \xi_{(0)}}{\partial t\partial u_{(0)}} +\frac{\partial^2 \tau_{(1)}}{\partial x\partial u_{(0)}} + \frac{\partial \xi_{(1)}}{\partial u_{(0)}}\right)\frac{\partial u_{(0)}}{ \partial t}\frac{\partial u_{(0)}}{ \partial x}\\
&+2\left(\frac{\partial^2 \tau_{(0)}}{\partial x\partial u_{(0)}}+\frac{\partial \xi_{(0)}}{\partial u_{(0)}}\right)\frac{\partial u_{(0)}}{ \partial t}\frac{\partial u_{(1)}}{ \partial x}+\left(\left(2\frac{\partial^3 \xi_{(0)}}{\partial x\partial u_{(0)}^2}-\frac{\partial^3 \eta_{(0)}}{\partial u_{(0)}^3}\right)u_{(1)}+2\frac{\partial^2 \xi_{(1)}}{\partial x\partial u_{(0)}} -\frac{\partial^2 \eta_{(1)}}{\partial u_{(0)}^2}\right)\left(\frac{\partial u_{(0)}}{ \partial x}\right)^2\\
&+2\left(\frac{\partial^2 \tau_{(0)}}{\partial x\partial u_{(0)}}+\frac{\partial \xi_{(0)}}{\partial u_{(0)}}\right)\frac{\partial u_{(0)}}{ \partial x}\frac{\partial u_{(1)}}{ \partial t}+2\left(2\frac{\partial^2 \xi_{(0)}}{\partial x\partial u_{(0)}}-\frac{\partial^2 \eta_{(0)}}{\partial u_{(0)}^2}\right)\frac{\partial u_{(0)}}{ \partial x}\frac{\partial u_{(1)}}{ \partial x}\\
&+\left(\left(\frac{\partial^3 \tau_{(0)}}{\partial x^2\partial u_{(0)}}-\frac{\partial^2 \tau_{(0)}}{\partial t\partial u_{(0)}}+2\frac{\partial^2 \xi_{(0)}}{\partial x\partial u_{(0)}}\right)u_{(1)}-\left(\frac{\partial^2 \tau_{(0)}}{\partial t^2}-2\frac{\partial^2 \eta_{(0)}}{\partial t\partial u_{(0)}}-\frac{\partial^2 \tau_{(1)}}{\partial t\partial x}+\frac{\partial \tau_{(1)}}{\partial t}-2\frac{\partial^2 \xi_{(1)}}{\partial x}\right)\right)\frac{\partial u_{(0)}}{\partial t}\\
&+\left(\left(\frac{\partial^3 \xi_{(0)}}{\partial x^2\partial u_{(0)}}-2\frac{\partial^3 \eta_{(0)}}{\partial x\partial u_{(0)}^2}-\frac{\partial^2 \xi_{(0)}}{\partial t\partial u_{(0)}}\right)u_{(1)}-\left(\frac{\partial^2 \xi_{(0)}}{\partial t^2}-\frac{\partial^2 \xi_{(1)}}{\partial x^2}+2\frac{\partial^2 \eta_{(1)}}{\partial x\partial u_{(0)}}+\frac{\partial \xi_{(1)}}{\partial t}\right)\right)\frac{\partial u_{(0)}}{\partial x}\\
&+\left(\frac{\partial^2 \tau_{(0)}}{\partial x^2}-\frac{\partial \tau_{(0)}}{\partial t}+2\frac{\partial \xi_{(0)}}{\partial x}\right)\frac{\partial u_{(1)}}{\partial t}+\left(\frac{\partial^2 \xi_{(0)}}{\partial x^2}-2\frac{\partial^2 \eta_{(0)}}{\partial x\partial u_{(0)}}-\frac{\partial \xi_{(0)}}{\partial t}\right)\frac{\partial u_{(1)}}{\partial x}\\
&\left.-\left(\frac{\partial^3 \eta_{(0)}}{\partial x^2\partial u_{(0)}}-\frac{\partial^2 \eta_{(0)}}{\partial t \partial u_{(0)}}\right)u_{(1)}+\frac{\partial^2 \eta_{(0)}}{\partial t^2}-\frac{\partial^2 \eta_{(1)}}{\partial x^2}+\frac{\partial \eta_{(1)}}{\partial t}\right)=O(\varepsilon^2).
\end{align*}
}
This relation turns out to be a polynomial in the unknowns $u_{(1)}$, $\displaystyle\frac{\partial u_{(0)}}{\partial t}$, $\displaystyle\frac{\partial u_{(0)}}{\partial x}$, $\displaystyle\frac{\partial u_{(1)}}{\partial t}$, $\displaystyle\frac{\partial u_{(1)}}{\partial x}$, $\displaystyle\frac{\partial^2 u_{(0)}}{\partial t^2}$, $\displaystyle\frac{\partial^2 u_{(0)}}{\partial t\partial x}$, $\displaystyle\frac{\partial u_{(1)}}{\partial t\partial x}$; by equating to zero the coefficients of these terms, we obtain an overdetermined linear system whose integration allows us to recover the following approximate infinitesimals:
\begin{equation}
\begin{aligned}
&\tau_{(0)}=\kappa_1,\qquad \xi_{(0)}=2\kappa_2 t+\kappa_3,\qquad \eta_{(0)}=-\kappa_2 xu_{(0)}+\kappa_4 u_{(0)}+f_1(t,x),\\
&\tau_{(1)}=4\kappa_5 t^2+2\kappa_6 t+2\kappa_2 x+\kappa_7,\qquad \xi_{(1)}=4\kappa_5 t x+2\kappa_8 t+\kappa_6 x+\kappa_9,\\
&\eta_{(1)}=-\kappa_5 (2t+x^2) u_{(0)}-\kappa_8 x u_{(0)}+\kappa_{10}u_{(0)}+f_2(t,x),
\end{aligned}
\end{equation}
where $\kappa_i$ $(i=1,\ldots,10)$ are constants and $f_i(t,x)$ ($i=1,2$) are functions of the indicated arguments, along with the constraints
\begin{equation}
\begin{aligned}
&\frac{\partial f_1(t,x)}{\partial t}-\frac{\partial^2 f_1(t,x)}{\partial x^2}=0,\\
&\frac{\partial^2 f_1(t,x)}{\partial t^2}-\frac{\partial^2 f_2(t,x)}{\partial x^2}+\frac{\partial f_2(t,x)}{\partial t}=0.
\end{aligned}
\end{equation}
\end{example}

\subsection{Approximate $Q$-conditional symmetries} 
It is known that many equations of interest in concrete applications possess poor Lie symmetries. More rich reductions leading to wide classes of exact solutions are possible by using the conditional symmetries, which are obtained by appending to the equations at hand some differential constraints. An important class is represented by Q-conditional symmetries, where the constraints to be added to the differential equation at hand are the invariant surface conditions and their differential consequences. 
It is easily proved that a (classical) Lie symmetry is a Q-conditional symmetry. However, differently from Lie symmetries, all possible conditional symmetries of a differential equation form a set which is not a Lie algebra in the general case.
Moreover, if the infinitesimals of a Q-conditional symmetry are multiplied by an arbitrary smooth function $f(\mathbf{x},\mathbf{u})$ we have still a Q-conditional symmetry. 

Remarkably, Q-conditional symmetries allow for symmetry reductions of differential equations and provide explicit solutions not always obtainable with classical symmetries.

Combining the approximate symmetry theory introduced in \cite{DSGO-lieapprox} with the approach of conditional symmetries, we can define approximate Q-conditional symmetries \cite{GorgoneOliveri-EJDE}. 

\begin{definition}[Approximate Q-conditional symmetries]
Given the differential equation
\begin{equation}
\Delta\equiv \sum_{k=0}^p\varepsilon^k\widetilde{\Delta}_{(k)}\left(\mathbf{x},\mathbf{u}_{(0)},\mathbf{u}^{(r)}_{(0)},
\ldots,\mathbf{u}_{(k)},\mathbf{u}^{(r)}_{(k)}\right)=O(\varepsilon^{p+1}),
\end{equation}
and the approximate invariant surface condition
\begin{equation}
\mathbf{Q}\equiv \sum_{k=0}^p\varepsilon^k\left(\sum_{i=1}^n\widetilde{\xi}_{(k)i}(\mathbf{x},\mathbf{u}_{(0)},\ldots,\mathbf{u}_{(k)})
\frac{\partial \mathbf{u}}{\partial x_i}-
\widetilde{\boldsymbol \eta}_{(k)}(\mathbf{x},\mathbf{u}_{(0)},\ldots,\mathbf{u}_{(k)})\right)=O(\varepsilon^{p+1}),
\end{equation}
the approximate Q-conditional symmetries of order $p$ are found by requiring
\begin{equation}
\left.\Xi^{(r)}(\Delta)\right|_{\mathcal{M}}=O(\varepsilon^{p+1}),
\end{equation}
where $\mathcal{M}$ is the manifold of the jet space defined by the system of equations
\begin{equation}
\Delta=O(\varepsilon^{p+1}), \qquad \mathbf{Q}=O(\varepsilon^{p+1}), \qquad \frac{\partial^{|s|} \mathbf{Q}}{\partial x_1^{s_1}\ldots\partial x_n^{s_n}}=O(\varepsilon^{p+1}),
\end{equation}
where $1\le |s|=s_1+\ldots +s_n \le r-1$.
\end{definition}

For simplicity, hereafter we will consider  the case $p=1$; for higher values of $p$ things are  similar but at the cost of an increased amount of computations.

Since the multiplication of the infinitesimals of a Q-conditional symmetry by an arbitrary smooth function $f(\mathbf{x},\mathbf{u})$ gives still a Q-conditional symmetry, we can look for approximate  Q-conditional symmetries in $n$ different situations, where $n$ is the number of independent variables, by assuming
\begin{itemize}
\item $\xi_{(0)1}=1$, $\xi_{(k)1}=0$ for $k=1,\ldots,p$, or
\item $\xi_{(0)i}=1$,  $\xi_{(k)i}=0$, $\xi_{(0)j}=\xi_{(k)j}=0$ for $1\le j <i \le n$ and $k=1,\ldots,p$.
\end{itemize}

As far as the example~\ref{ex:simple} is concerned, if we want to determine the first order approximate Q-conditional symmetries, we have to insert in the invariance condition therein displayed the constraints
\[
\begin{aligned}
&\tau_{(0)}=1, \qquad  \tau_{(1)}=0,\\
&Q\equiv \frac{\partial u_{(0)}}{\partial t}+\xi_{(0)}\frac{\partial u_{(0)}}{\partial x}-\eta_{(0)}\\
&\quad +\varepsilon\left(
\frac{\partial u_{(1)}}{\partial t}+\xi_{(0)}\frac{\partial u_{(1)}}{\partial x}+\left(\xi_{(1)}+\frac{\partial\xi_{(0)}}{\partial u_{(0)}}u_{(1)}\right)\frac{\partial u_{(0)}}{\partial x}-\left(\eta_{(1)}+\frac{\partial\eta_{(0)}}{\partial u_{(0)}}u_{(1)}\right)\right)=O(\varepsilon^2),\\
&\frac{\partial Q}{\partial t}=\frac{\partial Q}{\partial x}=O(\varepsilon^2).
\end{aligned}
\]

As a consequence, we obtain a very long invariance condition (we omit to write it) from which we can extract the determining equations (that now are \emph{nonlinear}) for the Q-conditional symmetries; it is easy to recognize that the use of a computer algebra package may help in order to perform the length and tedious calculations.
 
\section{Application}
\label{sec3}
In this Section, we consider the equation
\begin{equation}
\label{RDChyp}
\varepsilon \frac{\partial^2 u}{\partial t^2}+\frac{\partial u}{\partial t}-\frac{\partial}{\partial x}\left(u\frac{\partial u}{\partial x}\right)-\alpha u \frac{\partial u}{\partial x}+\beta u(1-\gamma u)=0,
\end{equation}
where $\varepsilon$ is a small constant parameter,
that represents a hyperbolic version of the reaction-diffusion-convection equation \cite{Murray}
\begin{equation}
\label{RDC}
\frac{\partial u}{\partial t}-\frac{\partial}{\partial x}\left(u\frac{\partial u}{\partial x}\right)-\alpha u \frac{\partial u}{\partial x}+\beta u(1-\gamma u)=0,
\end{equation}
$\alpha$, $\beta$ and $\gamma$ being real positive constants. Equation \eqref{RDC} can model  a population moving to domains with lower density faster than the overpopulation is reached and contains a logistic reaction term; it also includes another transport phenomenon, namely convection. The introduction of a term with a second order time derivative provides the hyperbolic equation
\eqref{RDChyp}. Conditional symmetries as well as some exact solutions of  equation \eqref{RDC} have been considered in \cite{Plyukhin}. A special class of approximate conditional symmetries of equation \eqref{RDChyp}, and some associated approximate solutions have been obtained in \cite{GorgoneOliveri-EJDE}.

It can be easily proved that equation~\eqref{RDChyp} possesses only the exact symmetries corresponding to space and time translations.  Also the admitted approximate symmetries of \eqref{RDChyp} reduce to the translation of independent variables. This means that either exact or approximate invariant solutions 
are only those depending on just one independent variable 
($t$ or $x$) as well as the ones describing travelling waves. More complex approximate symmetry reductions can be obtained by looking for approximate Q-conditional symmetries.  We observe that the zeroth terms of approximate Q-conditional symmetries of \eqref{RDChyp} provide particular exact Q-conditional symmetries of \eqref{RDC}.

Expanding $u(t,x;\varepsilon)$ at first order in $\varepsilon$, 
\begin{equation}\label{expansion}
u(t,x;\varepsilon)=u_{(0)}(t,x)+\varepsilon u_{(1)}(t,x)+O(\varepsilon^2),
\end{equation}
we consider the approximate Q-conditional symmetries of \eqref{RDChyp} corresponding to the approximate generator \cite{DSGO-lieapprox}
\begin{equation}
\begin{aligned}
\Xi&\approx \frac{\partial}{\partial t}
+\left(\xi_{(0)}(t,x,u_{(0)})+\varepsilon\left(\xi_{(1)}(t,x,u_{(0)})+\frac{\partial \xi_{(0)}(t,x,u_{(0)})}{\partial u_{(0)}}u_{(1)}\right)\right)\frac{\partial}{\partial x}\\
&+\left(\eta_{(0)}(t,x,u_{(0)})+\varepsilon\left(\eta_{(1)}(t,x,u_{(0)})+\frac{\partial \eta_{(0)}(t,x,u_{(0)})}{\partial u_{(0)}}u_{(1)}\right)\right)\frac{\partial}{\partial u}.
\end{aligned}
\end{equation}
By using the approach described in \cite{DSGO-lieapprox}, a set of determining equations is obtained, and various approximate Q-conditional symmetries can be recovered. 
In what follows, we limit ourselves to list those leading to representations of the approximate solutions which can be defined for all $t$. All the required computations have been done with the help of the program ReLie \cite{Oliveri:relie} written in the computer algebra system Reduce \cite{Reduce}.

The representations of the approximate solutions have to satisfy the approximate invariant surface condition; by separating terms at each order of $\varepsilon$, we have: 
\begin{equation}\label{AISC}
\begin{aligned}
&\frac{\partial u_{(0)}}{\partial t}+\xi_{(0)}\frac{\partial u_{(0)}}{\partial x}-\eta_{(0)}=0,\\
&\frac{\partial u_{(1)}}{\partial t}+\xi_{(0)}\frac{\partial u_{(1)}}{\partial x}+\xi_{(1)}\frac{\partial u_{(0)}}{\partial x}+\frac{\partial \xi_{(0)}}{\partial u_{(0)}}\frac{\partial u_{(0)}}{\partial x}u_{(1)}-\frac{\partial \eta_{(0)}}{\partial u_{(0)}}u_{(1)}-\eta_{(1)}=0;
\end{aligned}
\end{equation}
then, by inserting the solutions of \eqref{AISC} into Equation \eqref{RDChyp}, and solving the associated reduced systems of ordinary differential equations, some approximate Q-conditional solutions, for every set of approximate generators, can be explicitly determined.

In order to classify the approximate Q-conditional symmetries admitted by  Eq. \eqref{RDChyp} and the corresponding approximate solutions, three cases need to be distinguished:
\begin{enumerate}
\item[(i)] $8\beta\gamma-\alpha^2=-\delta^2$;
\item[(ii)] $8\beta\gamma-\alpha^2=0$;
\item[(iii)] $8\beta\gamma-\alpha^2=\delta^2$.
\end{enumerate}
For each case, because the determining equations are nonlinear, various sets of Q-conditional approximate
symmetries are recovered, as shown below, where the generators of Q-conditional approximate symmetries we found, the consequent representations of the approximate solutions, and the result of the integration of the reduced ordinary differential equations are listed. Often we are able to give the most general solution of the reduced equations. Some of these approximate solutions,
with suitable choices of the involved parameters, are also plotted. 
In passing, we observe that for $\varepsilon=0$ the obtained solutions satisfy the
unperturbed parabolic reaction-diffusion-convection Equation \eqref{RDC}.

\subsection{The case (i): $8\beta\gamma-\alpha^2=-\delta^2$}
\ \\
In this case, six different sets of Q-conditional approximate symmetries of \eqref{RDChyp} have been determined.

Using the generators
\begin{equation}
\begin{aligned}
\xi_{(0)}&=0,\qquad \eta_{(0)}=-\frac{\beta\kappa_1 \exp(\beta t)}{\kappa_1 \exp(\beta t)+1}u_{(0)},\qquad
\xi_{(1)}=\frac{4\kappa_2}{\kappa_1\exp(\beta t)+1},\\
\eta_{(1)}&=\frac{\exp(-\beta t-\frac{\alpha+\delta}{2}x)}{(\kappa_1\exp(\beta t)+1)^2 }\frac{\kappa_3\exp(\delta x)+\kappa_4}{u_{(0)}}-\frac{\exp(\beta t)}{(\kappa_1\exp(\beta t)+1)^2}\times\\
&\times\left(2\beta^2\kappa_1\log(\kappa_1\exp(\beta t)+1)
+\beta^2 \kappa_1(\kappa_1\exp(\beta t)-\beta t)+\kappa_1\kappa_2\alpha-\kappa_5\right)u_{(0)},
\end{aligned}
\end{equation}
and integrating the approximate invariant surface conditions \eqref{AISC}, we get the following representation of the solution:
\begin{equation}
\begin{aligned}
 u_{(0)}(t,x)&=\frac{U_0(x)}{\kappa_1\exp(\beta t)+1},\\
 u_{(1)}(t,x)&=\frac{1}{\kappa_1\exp(\beta t)+1}\left(-\frac{\exp\left(-\beta t-\frac{\alpha+\delta}{2}x\right)}{\beta}
\frac{\kappa_3\exp(\delta x)+\kappa_4}{U_0(x)}\right.\\
&-\frac{\beta^2\kappa_1(\kappa_1\exp(\beta t)(2\log(\kappa_1\exp(\beta t)+1)-\beta t)-1)-\kappa_1\kappa_2\alpha+\kappa_5}{\beta\kappa_1(\kappa_1\exp(\beta t)+1)}U_0(x)\\
&+4\left.\frac{\kappa_2}{\beta}(\log(\kappa_1\exp(\beta t)+1)-\beta t)U_0^\prime(x)+U_1(x)\right),
\end{aligned}
\end{equation}
where $U_0(x)$ and $U_1(x)$ are unknown functions to be determined. 

\begin{figure}
\begin{center}
\includegraphics[width=0.6\textwidth]{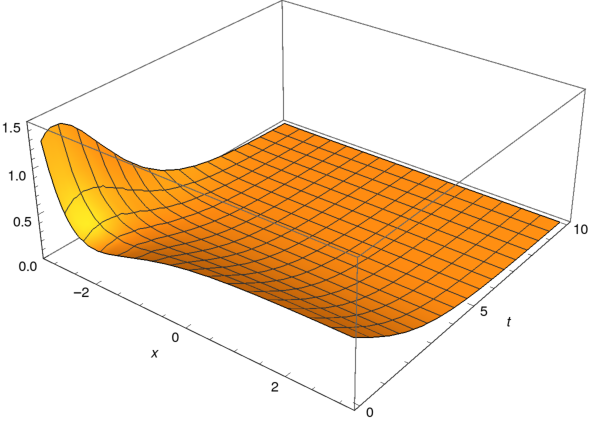}
\end{center}
\caption{\label{fig1a}Plot of solution \eqref{sol1a} with the parameters $\varepsilon= 0.03$, $\alpha=2$, 
$\beta=0.65$, $\gamma=0.77$, $c_1= 1$, $c_2= 1$, $\kappa_1=1$, $\kappa_2=2$, $\kappa_3=1$, $\kappa_4=2$, $\kappa_5=5.42$.}
\end{figure}

Now, by inserting this solution into the system \eqref{RDChyp}, and solving the corresponding reduced system, the associated approximate Q-conditional solution has the form:
\begin{equation}
\label{sol1a}
\begin{aligned}
 u(t,x;\varepsilon)&=\frac{8\beta}{(\alpha^2-\delta^2)(\kappa_1\exp(\beta t)+1)}\\
 &+\varepsilon\left(-\frac{\alpha^2-\delta^2}{8\beta^2}\exp\left(-\beta t-\frac{\alpha+\delta}{2}x\right)
(\kappa_3\exp(\delta x)+\kappa_4)\right.\\
 &-\frac{8\exp(\beta t)\left(\beta^2\kappa_1(2\log(\kappa_1\exp(\beta t)+1)-\beta t+1)+\kappa_1\kappa_2\alpha-\kappa_5\right)}{(\alpha^2-\delta^2)(\kappa_1\exp(\beta t)+1)^2}\\
 &+\left.\frac{\exp\left(-\left(\alpha+\theta\right)x/2\right)\left(c_1\exp\left(\theta x\right)+c_2\right)}{\kappa_1\exp(\beta t)+1}\right),
\end{aligned}
\end{equation}
where $\displaystyle\theta=\sqrt{\frac{\alpha^2+\delta^2}{2}}$, whereas $c_1$ and $c_2$ are arbitrary constants. A plot of this solution is displayed in Figure~\ref{fig1a}.

\begin{figure}
\begin{center}
\includegraphics[width=0.6\textwidth]{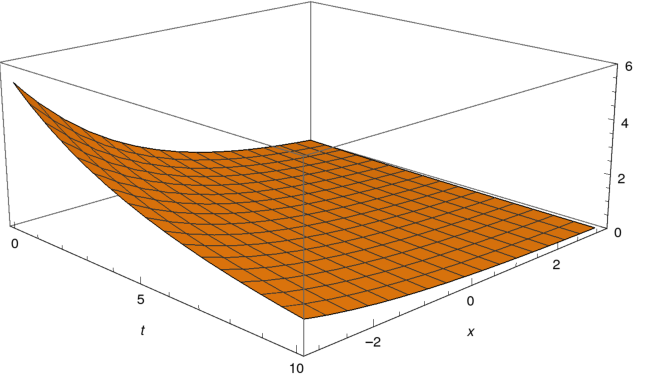}
\end{center}
\caption{\label{fig1b}Plot of solution \eqref{sol1b} with the parameters $\varepsilon= 0.03$, $\alpha=2$, 
$\beta=0.15$, $\gamma=3.31$, $c_1= 1$, $c_2= 0.01$, $c_3=-1$, $c_4=0$, $\kappa_1=1$, $\kappa_2=2$, $\kappa_3=1$, $\kappa_4=2$, $\kappa_5=5.02$, $\kappa_6=0.4$, $\kappa_7=0.9$, $\kappa_8=1$.}
\end{figure}

By integrating the approximate invariant surface conditions \eqref{AISC}, along with
\begin{equation}
\begin{aligned}
 \xi_{(0)}&=0,\qquad \eta_{(0)}=-\beta u_{(0)},\\
 \xi_{(1)}&=4\exp(-\beta t-\delta x)\left(\kappa_{1}\exp(2\delta x)+\kappa_{2}\exp(\delta x)+\kappa_3\right)\\
 &+4\exp\left(\beta t+\frac{\alpha-\delta}{2}x\right)(\kappa_4\exp(\delta x)+\kappa_{5})u_{(0)}^2,\\
 \eta_{(1)}&=\exp\left(-3\beta t-\frac{\alpha+\delta}{2}x\right)\frac{\kappa_{6}\exp(\delta x)+\kappa_{7}}{u_{(0)}}\\
 &-\exp(-\beta t-\delta x)\left(\kappa_{1}(\alpha-\delta)\exp(2\delta  x)+(\kappa_{2}\alpha+\kappa_8)\exp(\delta x)+\kappa_3(\alpha+\delta)\right)u_{(0)}\\
&-\beta^2 u_{(0)}
-\exp\left(\beta t+\frac{\alpha-\delta}{2} x\right)\left(\kappa_4(\alpha-\delta)\exp(\delta x)+\kappa_{5}(\alpha+\delta)\right)u_{(0)}^3,
\end{aligned}
\end{equation}
we get
\begin{equation}
\begin{aligned}
 u_{(0)}(t,x)&=\exp(-\beta t)U_0(x),\\
 u_{(1)}(t,x)&=-\frac{\exp\left(-2\beta t-\frac{\alpha+\delta}{2}x\right)}{\beta}\frac{\kappa_{6}\exp(\delta x)+\kappa_{7}}{U_0(x)}\\
 &+\frac{\exp(-2\beta t)}{\beta}\left(\kappa_{1}(\alpha-\delta)\exp(\delta x)+\kappa_{3}(\alpha+\delta)\exp(-\delta x)\right.\\
 &+\kappa_2\alpha+\kappa_8-\left.\exp(\beta t)\beta^3 t\right)U_0(x)+\exp(-\beta t)U_1(x)\\
 &+\frac{\exp\left(-2\beta t+\frac{\alpha-\delta}{2}x\right)}{\beta}\left(\kappa_4(\alpha-\delta)\exp(\delta x)+\kappa_5(\alpha+\delta)\right)U_0^3(x)\\
 &+4\frac{\exp(-2\beta t-\delta x)}{\beta}\left(\kappa_1\exp(2\delta x)+\kappa_{2}\exp(\delta x)+\kappa_{3}\right)U_0^\prime(x)\\
 &+4\frac{\exp\left(-2\beta t+\frac{\alpha-\delta}{2}x\right)}{\beta}(\kappa_4\exp(\delta x)+\kappa_5)U_0^2(x) U_0^\prime(x),
\end{aligned}
\end{equation}
and the corresponding approximate Q-conditional solution is written as
\begin{equation}
\label{sol1b}
\begin{aligned}
 u(t,x;\varepsilon)&=c_1\exp\left(-\beta t-\frac{\alpha+\delta}{4}x\right)\sqrt{\exp(\delta x)+c_2} \\
 &+\varepsilon\left(\frac{\exp(-\beta t)}{c_1^2\sqrt{c_2}\delta\sqrt{\exp(\delta x)+c_2}}\left(-\frac{4}{(\alpha-3\delta)(\alpha-\delta)}\right.\right.\times \\
 &\times\left(2c_1^2c_2^2\delta(\kappa_1(\alpha-3\delta)-2c_1^2\kappa_4\delta)+(\alpha-\delta)(2c_1^2\kappa_3\delta-\kappa_7)\right. \\
 &+\left.\phantom{\frac{}{}}c_2(2c_1^2\delta(2(\kappa_2+c_1^2\kappa_5)\delta-\kappa_2\alpha+\kappa_8)+\kappa_6(\alpha-3\delta))\right)\times \\
 &\times {_2F_1}\left(\frac{1}{2},\frac{\alpha-3\delta}{4\delta},\frac{\alpha+\delta}{4\delta};-\frac{\exp(\delta x)}{c_2}\right)-\frac{4}{\alpha+\delta}\times \\
 &\times\left(c_1^2((c_2(2c_1^2(c_2\kappa_4-\kappa_5)+\kappa_2)-2\kappa_3)\delta-c_2\kappa_8)+\kappa_7\right)\times \\
 &\times {_2F_1}\left(\frac{1}{2},\frac{\alpha+\delta}{4\delta},\frac{\alpha+5\delta}{4\delta};-\frac{\exp(\delta x)}{c_2}\right)-4\frac{\exp(\delta x)}{\alpha+5\delta}\times \\
 &\times\left(c_1^2((2c_1^2(c_2\kappa_4-\kappa_5)-\kappa_2+2c_2\kappa_1)\delta-\kappa_8)+\kappa_6\right)\times \\
 &\times \left.{_2F_1}\left(\frac{1}{2},\frac{\alpha+5\delta}{4\delta},\frac{\alpha+9\delta}{4\delta};-\frac{\exp(\delta x)}{c_2}\right)\right)
+\frac{\exp\left(-2\beta t-\frac{\alpha+\delta}{4}x\right)}{c_1\beta\sqrt{\exp(\delta x)+c_2}}\times \\
 &\times\left(-\exp(\delta x)\left(c_1^2((2c_1^2(c_2\kappa_4-\kappa_5)+2c_2\kappa_1-\kappa_2)\delta-\kappa_8)+\kappa_6\right)\right. \\
 &-\left.c_1^2((c_2(2c_1^2(c_2\kappa_4-\kappa_5)+\kappa_2)-2\kappa_3)\delta-c_2\kappa_8)-\kappa_7\right) \\
 &+\frac{\exp\left(-\beta t-\frac{\alpha+\delta}{4}x\right)}{c_1^2\delta(\alpha-\delta)(\exp(\delta x)+c_2)}\left(c_1^2(\alpha-\delta)((c_4-c_1\delta\beta^2t)\exp(\delta x)\right. \\
 &+(c_3-c_1c_2\beta^2 t)\delta)\sqrt{\exp(\delta x)+c_2}+4\exp\left(\frac{\alpha+5\delta}{4}x\right)\times \\
 &\times\left.\left(c_1^2((2c_1^2(c_2\kappa_4-\kappa_5)+2c_2\kappa_1-\kappa_2)\delta-\kappa_8)+\kappa_6\right)\right)\\
 &+\frac{4c_2\exp(-\beta t)}{\exp(\delta x)+c_2}\left.
\frac{c_1^2((2(c_1^2(c_2\kappa_4-\kappa_5)+c_2\kappa_1)-\kappa_2)\delta-\kappa_8)+\kappa_6}{c_1^2\delta(\alpha-\delta)}
\right),
\end{aligned}
\end{equation}
${_2F_1}$ being the \textit{hypergeometric function} defined as
\begin{equation}
{_2F_1}(a,b,c;x)=\sum_{k=0}^{\infty}\frac{(a)_k(b)_k}{(c)_k}\frac{x^k}{k!},
\end{equation}
where
\[
(a)_0=1,\quad (a)_k=\prod_{i=0}^{k-1} (i+a).
\]
A plot of this solution, fixing the values of the parameters therein involved, is displayed in Figure~\ref{fig1b}.

Taking
\begin{equation}
\begin{aligned}
\xi_{(0)}&=-\frac{4\beta}{\alpha+\delta},\qquad \eta_{(0)}=0,\\
\xi_{(1)}&=4\left(\kappa_1\exp(\beta t)+\kappa_2\exp\left(\frac{\beta(\alpha-7\delta)}{\alpha+\delta}t-\delta x\right)+\kappa_3\right),\\
\eta_{(1)}&=-(\alpha+\delta)\exp\left(\frac{(\alpha^2-8\alpha\delta-\delta^2)\beta t-\alpha^2 \delta x}{\alpha^2-\delta^2}\right)\times\\
&\times\left(\kappa_1 \exp\left(\frac{\alpha\delta(8\beta t+\alpha x)}{\alpha^2-\delta^2}\right)+\kappa_2\exp\left(\frac{\delta^2(8\beta t+\delta x)}{\alpha^2-\delta^2}\right)\right)u_{(0)},
\end{aligned}
\end{equation}
the integration of the conditions \eqref{AISC} provides the following representation of the solution:
\begin{equation}
\begin{aligned}
 u_{(0)}(t,x)&=U_0(\omega),\\
 u_{(1)}(t,x)&=-\left(\frac{\kappa_1}{\beta}\exp(\beta t)+\frac{\alpha+\delta}{\beta(\alpha-3\delta)}\kappa_2\exp\left(\frac{\beta(\alpha-3\delta)}{\alpha+\delta}t-\delta\omega\right)\right)\times\\
 &\times \left((\alpha+\delta)U_0(\omega)+4U_0^\prime(\omega)\right)-4\kappa_3 t U_0^\prime(\omega)+U_1(\omega),
\end{aligned}
\end{equation}
with $\omega=\displaystyle\frac{4\beta}{\alpha+\delta}t+x$.

\begin{figure}
\begin{center}
\includegraphics[width=0.6\textwidth]{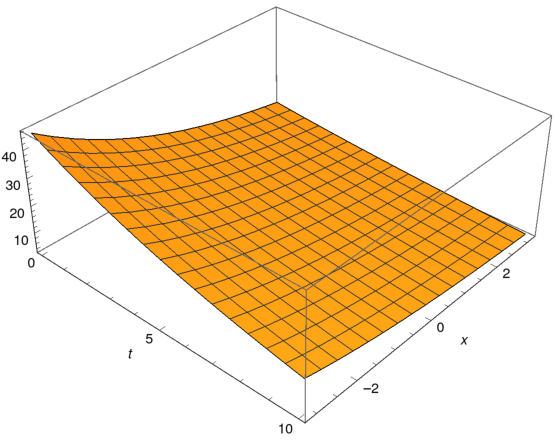}
\end{center}
\caption{\label{fig1c}Plot of solution \eqref{sol1c} with the parameters $\varepsilon= 0.03$, $\alpha=0.7$, 
$\beta=0.1$, $\gamma=0.53$, $c_1= 1$, $c_2= 0$, $\kappa_1=1$, $\kappa_2=2$.}
\end{figure}

The associated approximate Q-conditional solution (a plot is in Figure~\ref{fig1c}) reads
\begin{equation}
\label{sol1c}
\begin{aligned}
 u(t,x;\varepsilon)&=\frac{8\beta}{\alpha^2-\delta^2}\\
 &+\varepsilon\left(-8\frac{(3\delta-\alpha)\kappa_1\exp(\beta t)-(\alpha+\delta)\kappa_2\exp\left(\frac{\beta(\alpha-7\delta)}{\alpha+\delta}t-\delta x\right)}{(3\delta-\alpha)(\alpha-\delta)}\right.\\
 &+\exp\left(-\frac{\alpha+\delta+\sqrt{(3\delta-\alpha)(\alpha+\delta)}}{4}\left(\frac{4\beta}{\alpha+\delta}t+x\right)\right)\times\\
 &\left.\times\left(c_1\exp\left(\frac{\sqrt{(3\delta-\alpha)(\alpha+\delta)}}{2}\left(\frac{4\beta}{\alpha+\delta}t+x\right)\right)+c_2\right)\right).
\end{aligned}
\end{equation}

Using the generators
\begin{equation}\label{gen112}
\begin{aligned}
&\xi_{(0)}=\kappa_1,\qquad \eta_{(0)}=0,\\
&\xi_{(1)}=\kappa_2\exp(\beta t)+\kappa_3,\qquad
\eta_{(1)}=\frac{\kappa_2}{\kappa_1}\beta\exp(\beta t)u_{(0)},
\end{aligned}
\end{equation}
with $\kappa_1 \neq 0,-\displaystyle\frac{4\beta}{\alpha+\delta},\frac{4\beta}{\alpha-\delta}$,
the integration of the conditions \eqref{AISC}, provides
\begin{equation}\label{rep112}
\begin{aligned}
 u_{(0)}(t,x)&=U_0(\omega),\\
 u_{(1)}(t,x)&=\frac{\kappa_2}{\kappa_1}\exp(\beta t)U_0(\omega)-\left(\frac{\kappa_2}{\beta}\exp(\beta t)+\kappa_3 t\right)U_0^\prime(\omega)+U_1(\omega),
\end{aligned}
\end{equation}
with $\omega=-\kappa_1 t+x$.
The corresponding approximate Q-conditional solution we find turns out to be
\begin{equation}
\begin{aligned}
 u(t,x;\varepsilon)&=\frac{8\beta}{\alpha^2-\delta^2}+\varepsilon\left(\frac{8\beta}{\alpha^2-\delta^2}\frac{\kappa_2}{\kappa_1}\exp(\beta t)+\exp\left(\left(\frac{\kappa_1(\alpha^2-\delta^2)}{8\beta}\right.\right.\right.\\
 &+\left.\left.\alpha+\sqrt{\frac{(\kappa_1(\alpha^2-\delta^2)+8\alpha\beta)^2}{64\beta^2}-\frac{\alpha^2-\delta^2}{2}}\right)\frac{\kappa_1 t-x}{2}\right)\times\\
 &\times\left.\left(c_1\exp\left(\sqrt{\frac{(\kappa_1(\alpha^2-\delta^2)+8\alpha\beta)^2}{64\beta^2}-\frac{\alpha^2-\delta^2}{2}}(-\kappa_1 t+x)\right)+c_2\right)\right).
\end{aligned}
\end{equation}

Moreover, with the approximate generators
\begin{equation}
\begin{aligned}
\xi_{(0)}&=-\frac{4\beta}{\alpha-\delta},\qquad \eta_{(0)}=0,\\
\xi_{(1)}&=4\left(\kappa_1\exp(\beta t)+\kappa_2\exp\left(\frac{\beta(\alpha+7\delta) }{\alpha-\delta}t+\delta x\right)+\kappa_3\right),\\
\eta_{(1)}&=-(\alpha-\delta)\exp\left(-\frac{\delta^2}{\alpha-\delta}x\right)\times\\
&\times\left(\kappa_1\exp\left(\beta t+\frac{\delta^2}{\alpha-\delta}x\right)+\kappa_2 \exp\left(\frac{\beta(\alpha+7\delta) t+\alpha\delta x}{\alpha-\delta}\right)\right)u_{(0)},
\end{aligned}
\end{equation}
we obtain the following representation of the solution:
\begin{equation}
\begin{aligned}
 u_{(0)}(t,x)&=U_0(\omega),\\
 u_{(1)}(t,x)&=\left(\frac{\kappa_1(\delta-\alpha)}{\beta}\exp(\beta t)\right.
-\left.\frac{\kappa_2(\alpha-\delta)^2}{\beta(\alpha+3\delta)}\exp\left(\frac{\beta(\alpha+3\delta)}{\alpha-\delta}t+\delta\omega\right)\right)U_0(\omega)\\
 &-4\left(\frac{\kappa_1}{\beta}\exp(\beta t)+\frac{\alpha-\delta}{\beta(\alpha+3\delta)}\kappa_2\exp\left(\frac{\beta(\alpha+3\delta)}{\alpha-\delta}t+\delta\omega\right)
+\kappa_3 t\right)U_0^\prime(\omega)\\
&+U_1(\omega),
\end{aligned}
\end{equation}
with $\omega=\displaystyle\frac{4\beta}{\alpha-\delta}t+x$.
An approximate Q-conditional solution is given by
\begin{equation}
\begin{aligned}
 &u(t,x;\varepsilon)=\frac{8\beta}{\alpha^2-\delta^2}
-8\varepsilon\frac{(3\delta+\alpha)\kappa_1\exp(\beta t)+(\alpha-\delta)\kappa_2\exp\left(\frac{\beta(\alpha+7\delta)}{\alpha-\delta}t+\delta x\right)}{(3\delta+\alpha)(\alpha+\delta)}.
\end{aligned}
\end{equation}

Finally, with the approximate generators
\begin{equation}
\begin{aligned}
\xi_{(0)}&=\beta\kappa_1\exp(-\beta t),\qquad \eta_{(0)}=-\beta u_{(0)},\\
\xi_{(1)}&=(\kappa_2-\kappa_1\beta^3 t)\exp(-\beta t)+\kappa_1(\kappa_3-\kappa_1\alpha\beta^2)\exp(-2\beta t)\\
&-\beta^2\frac{\alpha^2-\delta^2}{8}\kappa_1^3\exp(-3\beta t),\\
 \eta_{(1)}&=-\beta^3 \kappa_1^2\exp(-2\beta t)
+\left(\frac{\beta^2(\alpha^2-\delta^2)}{4}\kappa_1^2\exp(-2\beta t)-\kappa_3\exp(-\beta t)-\beta^2\right)u_{(0)},
\end{aligned}
\end{equation}
we have the representation of the solution in the form
\begin{equation}\label{rep115}
\begin{aligned}
 u_{(0)}(t,x)&=\exp(-\beta t)U_0(\omega),\\
 u_{(1)}(t,x)&=-\frac{\exp(-3\beta t)}{\beta}\left(\beta^2\kappa_1^2\frac{\alpha^2-\delta^2}{8}-\kappa_3\exp(\beta t)+\exp(2\beta t)\beta^3 t\right)U_0(\omega)\\
 &-\frac{\exp(-4\beta t)}{\beta}\left(\phantom{\frac{}{}}\exp(2\beta t)(\beta^2\kappa_1(\beta t+1)-\kappa_2)\right.\\
 &+\left.\frac{\kappa_1}{2}(\kappa_1\alpha\beta^2-\kappa_3)\exp(\beta t)+\beta^2\kappa_1^3\frac{\alpha^2-\delta^2}{24}\right)U_0^\prime(\omega)\\
 &+\beta^2\kappa_1^2\exp(-2\beta t)+\exp(-\beta t)U_1(\omega),
\end{aligned}
\end{equation}
with $\omega=\kappa_1\exp(-\beta t)+x$.
Such a representation does not allow us to solve in general the corresponding reduced system of ordinary differential equations; 
however, by choosing $\kappa_1=0$, we are able to get the following approximate Q-conditional solution:
\begin{equation}
\begin{aligned}
 u(t,x;\varepsilon)&=c_1\exp\left(-\beta t-\frac{\alpha+\delta}{4}x\right)\sqrt{\exp(\delta x)+c_2}\\
 &+\varepsilon\left(\frac{\sqrt{c_2}\exp(-\beta t)}{\sqrt{\exp(\delta x)+c_2}}\times\right.\\
 &\left(\frac{4(\kappa_2(\alpha-\delta)-2\kappa_3)}{\alpha^2-4\alpha\delta+3\delta^3} {_2F_1}\left(\frac{1}{2},\frac{\alpha-3\delta}{4\delta},\frac{\alpha+\delta}{4\delta};-\frac{\exp(\delta x)}{c_2}\right)\right.\\
 &-\frac{\kappa_2(\alpha+\delta)-4\kappa_3}{\delta(\alpha+\delta)} {_2F_1}\left(\frac{1}{2},\frac{\alpha+\delta}{4\delta},\frac{\alpha+5\delta}{4\delta};-\frac{\exp(\delta x)}{c_2}\right)\\
 &-\frac{\exp(\delta x)(\kappa_2(\alpha-\delta)-4\kappa_3)}{c_2\delta(\alpha+5\delta)} \left.{_2F_1}\left(\frac{1}{2},\frac{\alpha+5\delta}{4\delta},\frac{\alpha+9\delta}{4\delta};-\frac{\exp(\delta x)}{c_2}\right)\right)\\
 &-c_1\frac{\exp\left(-2\beta t-\frac{\alpha+\delta}{4}x\right)}{4\beta\sqrt{\exp(\delta x)+c_2}}\left(\exp(\delta x)(\kappa_2(\alpha-\delta)-4\kappa_3)+c_2(\kappa_2(\alpha+\delta)-4\kappa_3)\right)\\
&-\frac{\exp\left(-\beta t-\frac{\alpha+\delta}{4}x\right)}{\delta\sqrt{\exp(\delta x)+c_2}}\left(\exp(\delta x)(c_1\delta\beta^2 t-c_4)+(c_1c_2\beta^2t-c_3)\delta\right)\\
 &+\left.\frac{\exp(-\beta t)}{\delta(\alpha-\delta)}(\kappa_2(\alpha-\delta)-4\kappa_3)\right).
\end{aligned}
\end{equation}
\subsection{The case (ii): $8\beta\gamma-\alpha^2=0$}
\ \\
In this case, we are able to recover five different sets of Q-conditional approximate symmetries of \eqref{RDChyp}.

\begin{figure}
\begin{center}
\includegraphics[width=0.6\textwidth]{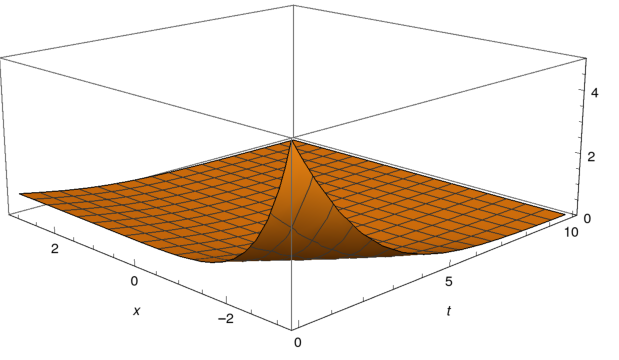}
\end{center}
\caption{\label{fig2a}Plot of solution \eqref{sol2a} with the parameters $\varepsilon= 0.03$, $\alpha=2$, 
$\beta=0.65$, $\gamma=0.77$, $c_1= 1$, $c_2= 1$, $\kappa_1=1$, $\kappa_2=0$, $\kappa_3=1$, $\kappa_4=2$, $\kappa_5=3$.}
\end{figure}

By using the approximate generators
\begin{equation}
\begin{aligned}
 \xi_{(0)}&=0,\qquad \eta_{(0)}=-\frac{\beta\kappa_1 \exp(\beta t)}{\kappa_1 \exp(\beta t)+1}u_{(0)},\\
 \xi_{(1)}&=\frac{4\kappa_2}{\kappa_1\exp(\beta t)+1},\\
 \eta_{(1)}&=\frac{\exp(-\beta t-\frac{\alpha}{2}x)}{(\kappa_1\exp(\beta t)+1)^2 }\frac{\kappa_3 x+\kappa_4}{u_{(0)}}-\frac{\exp(\beta t)}{(\kappa_1\exp(\beta t)+1)^2}\left(\phantom{\frac{}{}}\beta^2\kappa_1(2\log(\kappa_1\exp(\beta t)+1)\right.\\
 &+\left.\phantom{\frac{}{}}\kappa_1\exp(\beta t)-\beta t)+\kappa_1\kappa_2\alpha-\kappa_5 \right)u_{(0)},
\end{aligned}
\end{equation}
and integrating the approximate invariant surface conditions \eqref{AISC}, we get:
\begin{equation}
\begin{aligned}
 u_{(0)}(t,x)&=\frac{U_0(x)}{\kappa_1\exp(\beta t)+1},\\
 u_{(1)}(t,x)&=\frac{1}{\kappa_1\exp(\beta t)+1}\left(-\frac{\exp\left(-\beta t-\frac{\alpha}{2}x\right)}{\beta}
\frac{\kappa_3 x+\kappa_4}{U_0(x)}\right.\\
 &-\frac{\beta^2\kappa_1(\kappa_1\exp(\beta t)(2\log(\kappa_1\exp(\beta t)+1)-\beta t)-1)-\kappa_1\kappa_2\alpha+\kappa_5}{\beta\kappa_1(\kappa_1\exp(\beta t)+1)}
 U_0(x)\\
 &+4\left.\frac{\kappa_2}{\beta}(\log(\kappa_1\exp(\beta t)+1)-\beta t)U_0^\prime(x)+U_1(x)\right).
\end{aligned}
\end{equation}
Solving the reduced ordinary differential equations, the  approximate Q-conditional solution has the form:
\begin{equation}
\label{sol2a}
\begin{aligned}
 u(t,x;\varepsilon)&=\frac{8\beta}{\alpha^2(\kappa_1\exp(\beta t)+1)}\\
 &+\varepsilon\left(-\frac{\alpha^2}{8\beta^2}\exp\left(-\beta t-\frac{\alpha}{2}x\right)
(\kappa_3 x+\kappa_4)\right.\\
 &-\frac{8\exp(\beta t) \left(\beta^2\kappa_1(2\log(\kappa_1\exp(\beta t)+1)-\beta t+1)+\kappa_1\kappa_2\alpha-\kappa_5\right)}{\alpha^2(\kappa_1\exp(\beta t)+1)^2}\\
 &+\left.\frac{\exp\left(-\left(1+\frac{\sqrt{2}}{2}\right)\frac{\alpha x}{2}\right)}{\kappa_1\exp(\beta t)+1}\left(c_1\exp\left(\frac{\sqrt{2}}{2} \alpha x\right)+c_2\right)\right),
\end{aligned}
\end{equation}
with $c_1$ and $c_2$ arbitrary constants (Figure~\ref{fig2a} shows a  plot of this solution).

\begin{figure}
\begin{center}
\includegraphics[width=0.6\textwidth]{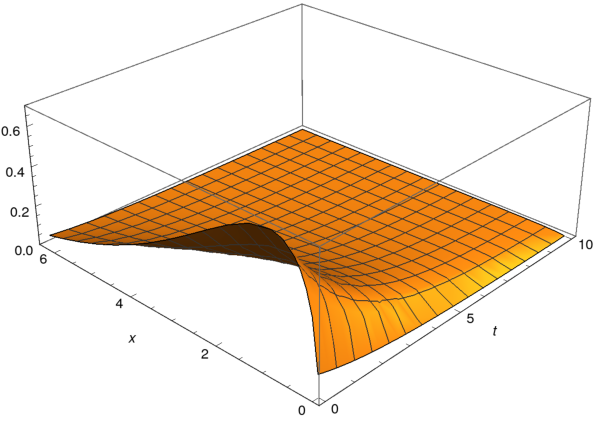}
\end{center}
\caption{\label{fig2b}Plot of solution \eqref{sol2b} with the parameters $\varepsilon= 0.03$, $\alpha=2$, 
$\beta=0.4$, $\gamma=1.25$, $c_1= 1$, $c_2= 0$, $c_3=0$, $\kappa_1=0$, $\kappa_2=1$, $\kappa_3=0$, $\kappa_4=0$, $\kappa_5=0$, $\kappa_6=0$, $\kappa_7=0$, $\kappa_8=0$.}
\end{figure}

Considering the approximate generators
\begin{equation}
\begin{aligned}
 \xi_{(0)}&=0,\qquad \eta_{(0)}=-\beta u_{(0)},\\
 \xi_{(1)}&=4\exp(-\beta t)\left(\kappa_1 x^2+\kappa_{2}x+\kappa_{3}\right)+4\exp\left(\beta t+\frac{\alpha}{2}x\right)(\kappa_4 x+\kappa_{5})u_{(0)}^2,\\
 \eta_{(1)}&=\exp\left(-3\beta t-\frac{\alpha}{2}x\right)\frac{\kappa_{6}x+\kappa_{7}}{u_{(0)}}\\
 &-\exp(-\beta t)(\kappa_{1}(\alpha x-2)x+\kappa_{2}(\alpha x -1)+\kappa_3\alpha+\kappa_8)u_{(0)}\\
 &-\beta^2 u_{(0)}-\exp\left(\beta t+\frac{\alpha}{2} x\right)(\kappa_4(\alpha x-2)+\kappa_5\alpha)u_{(0)}^3,
\end{aligned}
\end{equation}
the integration of the approximate invariant surface conditions \eqref{AISC} provides
\begin{equation}
\begin{aligned}
 u_{(0)}(t,x)&=\exp(-\beta t)U_0(x),\\
 u_{(1)}(t,x)&=-\frac{\exp\left(-2\beta t-\frac{\alpha}{2}x\right)(\kappa_{6}x+\kappa_{7})}{\beta U_0(x)}  \\
 &+\frac{\exp(-2\beta t)}{\beta}\left(\kappa_{1}(\alpha x-2)x+\kappa_{2}(\alpha x -1)+\kappa_3\alpha+\kappa_8\right.\\
 &-\left.\exp(\beta t)\beta^3 t\right)U_0(x)+\frac{\exp\left(-2\beta t+\frac{\alpha x}{2}\right)}{\beta}(\kappa_4(\alpha x-2)+\kappa_5\alpha)U_0^3(x)\\
 &+4\frac{\exp(-2\beta t)}{\beta}\left(\kappa_1 x^2+\kappa_{2}x+\kappa_{3}\right)U_0^\prime(x)\\
 &+4\frac{\exp\left(-2\beta t+\frac{\alpha x}{2}\right)}{\beta}(\kappa_4 x+\kappa_5)U_0^2(x) U_0^\prime(x)+\exp(-\beta t)U_1(x),
\end{aligned}
\end{equation}
and the corresponding approximate Q-conditional solution  (a plot is shown in Figure~\ref{fig2b}) 
turns out to be
\begin{equation}
\label{sol2b}
\begin{aligned}
 u(t,x;\varepsilon)&=c_1\exp\left(-\beta t-\frac{\alpha x}{4}\right)\sqrt{x}\\
 &+\varepsilon\left(\frac{\exp\left(-2\beta t-{\alpha x}/{4}\right)}{c_1\beta\sqrt{x}}\left((c_1^2(\kappa_2+2 c_1^2\kappa_5+\kappa_8)-\kappa_6)x+2 c_1^2\kappa_3-\kappa_7\right)\right.\\
 &+\exp\left(-\beta t-\frac{\alpha x}{4}\right)\left(\frac{(c_3-c_1\beta^2 t)x+c_2}{\sqrt{x}}\right.\\
 &+\frac{2\sqrt{\pi}}{c_1^2\alpha^2\sqrt{\alpha}}\left((2c_1^2(\kappa_2-\kappa_3\alpha+2 c_1^2\kappa_5+\kappa_8)-2\kappa_6+\kappa_7\alpha)\alpha x\right.\\
 &+\left.\left.4c_1^2(3(\kappa_2+2 c_1^2\kappa_5+\kappa_8)-\kappa_3\alpha)-12\kappa_6+2\kappa_7\alpha\right) \frac{\hbox{erfi}\left({\sqrt{\alpha x}}/{2}\right)}{\sqrt{x}}\right)\\
 &+\left.\frac{4\exp(-\beta t)}{c_1^2\alpha^2}\left(2c_1^2(\kappa_3\alpha-3(\kappa_2+2 c_1^2\kappa_5+\kappa_8))+6\kappa_6-\kappa_7\alpha\right)\right),
\end{aligned}
\end{equation}
where $\displaystyle\hbox{erfi}\left(\frac{\sqrt{\alpha x}}{2}\right)$ is the \textit{imaginary error function}, defined as
\begin{equation}
\mathrm{erfi}(x)=\frac{2}{\sqrt{\pi}}\int_{0}^{x}\exp\left(t^2\right) dt.
\end{equation}

\begin{figure}
\begin{center}
\includegraphics[width=0.6\textwidth]{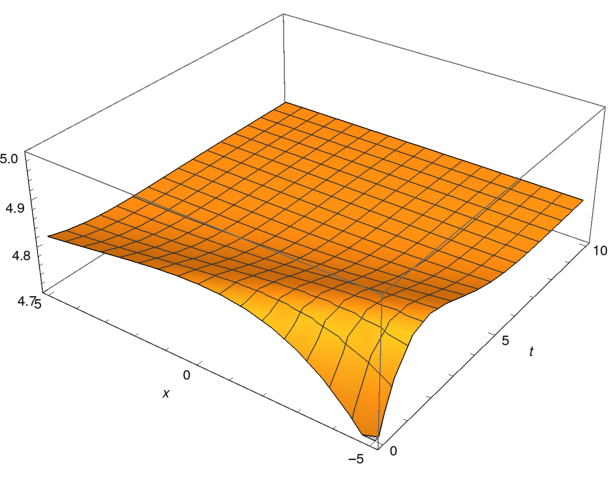}
\end{center}
\caption{\label{fig2c}Plot of solution \eqref{sol2c} with the parameters $\varepsilon= 0.03$, $\alpha=1$, 
$\beta=0.6$, $\gamma=0.21$, $c_1= 2$, $c_2= 2$, $\kappa_1=0$, $\kappa_2=0$.}
\end{figure}

Using the approximate generators
\begin{equation}
\begin{aligned}
\xi_{(0)}&=-\frac{4\beta}{\alpha},\qquad \eta_{(0)}=0,\\
\xi_{(1)}&=4\left(\exp(\beta t)(\kappa_1\alpha+\kappa_2\beta(8\beta t +\alpha x-8))+\kappa_3\right),\\
\eta_{(1)}&=-\alpha\exp(\beta t)(\kappa_1\alpha+\kappa_2\beta(8\beta t +\alpha x-8))u_{(0)},
\end{aligned}
\end{equation}
and integrating the conditions \eqref{AISC}, we get the following representation of the solution:
\begin{equation}
\begin{aligned}
u_{(0)}(t,x)&=U_0(\omega),\\
u_{(1)}(t,x)&=-\exp(\beta t)\left(\kappa_1\frac{\alpha}{\beta}+\kappa_2(4\beta t+\alpha\omega-12)\right)\times\\
 &\times \left(\alpha U_0(\omega)+4U_0^\prime(\omega)\right)-4\kappa_3 t U_0^\prime(\omega)+U_1(\omega),
\end{aligned}
\end{equation}
with $\omega=\displaystyle\frac{4\beta}{\alpha}t+x$.
The associated approximate Q-conditional solution (a plot is in Figure~\ref{fig2c}) reads
\begin{equation}
\label{sol2c}
\begin{aligned}
 u(t,x;\varepsilon)&=\frac{8\beta}{\alpha^2}+\varepsilon\left(-8\exp(\beta t)\left(\kappa_1+\kappa_2\frac{\beta}{\alpha}(8\beta t+\alpha x-12)\right)\right.\\
 &+\left.\exp\left(-\beta t -\frac{\alpha x}{4}\right)\left(c_1\sin\left(\beta t+\frac{\alpha x}{4}\right)+c_2\cos\left(\beta t+\frac{\alpha x}{4}\right)\right)\right).
\end{aligned}
\end{equation}

Another set of the approximate generators writes as \eqref{gen112}, and the corresponding approximate Q-conditional solution 
is given by
\begin{equation}
\begin{aligned}
 u(t,x;\varepsilon)&=
\frac{8\beta}{\alpha^2}+\varepsilon\left(\frac{8\beta}{\alpha^2}\frac{\kappa_2}{\kappa_1}\exp(\beta t)\right.\\
 &+\exp\left(\left(\frac{\kappa_1\alpha^2}{8\beta^2}+\alpha+\sqrt{\frac{\alpha^2(\kappa_1\alpha+8\beta)^2}{64\beta^2}-\frac{\alpha^2}{2}}\right)\frac{\kappa_1 t-x}{2}\right)\times\\
 &\times\left. \left(c_1\exp\left(\sqrt{\frac{\alpha^2(\kappa_1\alpha+8\beta)^2}{64\beta^2}-\frac{\alpha^2}{2}}(-\kappa_1 t+x)\right)+c_2\right)\right).
\end{aligned}
\end{equation}

Finally, the approximate generators
\begin{equation}
\begin{aligned}
 \xi_{(0)}&=\beta\kappa_1\exp(-\beta t),\qquad \eta_{(0)}=-\beta u_{(0)},\\
 \xi_{(1)}&=(\kappa_2-\kappa_1\beta^3 t)\exp(-\beta t)+\kappa_1(\kappa_3-\kappa_1\alpha\beta^2)\exp(-2\beta t)
-\kappa_1^3\frac{\alpha^2\beta^2}{8}\exp(-3\beta t),\\
 \eta_{(1)}&=-\beta^3\kappa_1^2\exp(-2\beta t)
+\left(\kappa_1^2\frac{\alpha^2\beta^2}{4}\exp(-2\beta t)-\kappa_3\exp(-\beta t)-\beta^2\right)u_{(0)}
\end{aligned}
\end{equation}
provide a representation of the solution obtainable from  \eqref{rep115} by setting $\delta=0$, \textit{i.e.},
\begin{equation}
\begin{aligned}
 u_{(0)}(t,x)&=\exp(-\beta t)U_0(\omega),\\
 u_{(1)}(t,x)&=-\frac{\exp(-3\beta t)}{\beta}\left(\kappa_1^2\frac{\alpha^2\beta^2}{8}-\kappa_3\exp(\beta t)+\exp(2\beta t)\beta^3 t\right)U_0(\omega)\\
 &-\frac{\exp(-4\beta t)}{\beta}\left(\phantom{\frac{}{}}\exp(2\beta t)(\beta^2\kappa_1(\beta t+1)-\kappa_2)\right.\\
 &+\left.\frac{\kappa_1}{2}(\kappa_1\alpha\beta^2-\kappa_3)\exp(\beta t)+\kappa_1^3\frac{\alpha^2\beta^2}{24}\right)U_0^\prime(\omega)\\
 &+\beta^2\kappa_1^2\exp(-2\beta t)+\exp(-\beta t)U_1(\omega),
\end{aligned}
\end{equation}
where $\omega=\kappa_1\exp(-\beta t)+x$.

\begin{figure}
\begin{center}
\includegraphics[width=0.6\textwidth]{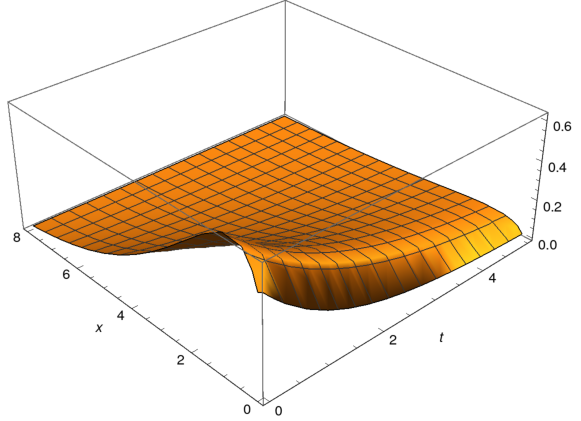}
\end{center}
\caption{\label{fig2d}Plot of solution \eqref{sol2d} with the parameters $\varepsilon= 0.03$, $\alpha=2$, 
$\beta=0.4$, $\gamma=1.25$, $c_1= 1$, $c_2= 0$, $c_3=0$, $\kappa_2=1$, $\kappa_3=1$.}
\end{figure}

Also in this case, we can not solve in general the corresponding reduced system; nevertheless, by taking $\kappa_1=0$, we obtain the following approximate Q-conditional solution:
\begin{equation}
\label{sol2d}
\begin{aligned}
 u(t,x;\varepsilon)&=c_1\exp\left(-\beta t-\frac{\alpha x}{4}\right)\sqrt{x} \\
 &+\varepsilon\left(\frac{c_1}{4\beta}\exp\left(-2\beta t-\frac{\alpha x}{4}\right)\frac{(4\kappa_3-\kappa_2\alpha)x+2\kappa_2}{\sqrt{x}}\right. \\
 &+\frac{\exp\left(-\beta t-\frac{\alpha x}{4}\right)}{\alpha^2}\left(\phantom{\frac{}{}}\alpha^2((c_2-c_1\beta^2t)x+c_3)\right. \\
 &+\left.\frac{2\sqrt{\pi}}{\sqrt{a}}\left((2\kappa_3-\kappa_2\alpha)\alpha x+4(3\kappa_3-\kappa_2\alpha)\right)\hbox{erfi}\left(\frac{\sqrt{\alpha x}}{2}\right)\right)\frac{1}{\sqrt{x}} \\
 &+\left.8\frac{\exp(-\beta t)}{\alpha^2}(\kappa_2\alpha-\kappa_3)\right).
\end{aligned}
\end{equation}
A plot of this solution is displayed in Figure~\ref{fig2d}.

\subsection{The case (iii): $8\beta\gamma-\alpha^2=\delta^2$}
\ \\
In this case, we obtain four different sets of Q-conditional approximate symmetries of \eqref{RDChyp}, but in only two cases
we obtain general real valued solutions.  

\begin{figure}
\begin{center}
\includegraphics[width=0.6\textwidth]{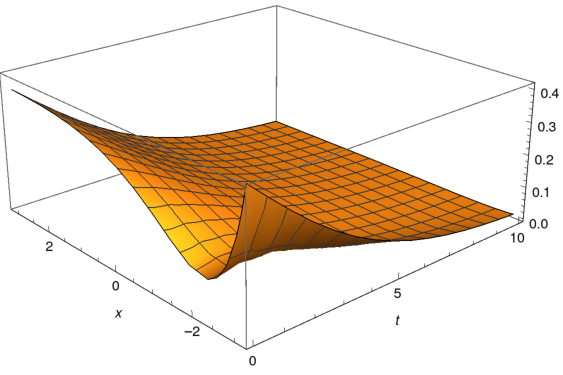}
\end{center}
\caption{\label{fig3a}Plot of solution \eqref{sol3a} with the parameters $\varepsilon= 0.03$, $\alpha=2$, 
$\beta=0.4$, $\gamma=1.33$, $c_1= 1$, $c_2= 0$, $\kappa_1=1$, $\kappa_2=0$, $\kappa_3=1$, $\kappa_4=1$, $\kappa_5=0$.}
\end{figure}

By using
\begin{equation}
\begin{aligned}
\xi_{(0)}&=0,\qquad \eta_{(0)}=-\frac{\beta\kappa_1 \exp(\beta t)}{\kappa_1 \exp(\beta t)+1}u_{(0)},\\
\xi_{(1)}&=\frac{4\kappa_2}{\kappa_1\exp(\beta t)+1},\\
\eta_{(1)}&=\frac{\exp(-\beta t-\frac{\alpha}{2}x)}{(\kappa_1\exp(\beta t)+1)^2 }\frac{\kappa_3\sin\left(\frac{\delta }{2}x\right)+\kappa_4\cos\left(\frac{\delta }{2}x\right)}{u_{(0)}}\\
&-\frac{\exp(\beta t)}{(\kappa_1\exp(\beta t)+1)^2}\left(\phantom{\frac{}{}}2\beta^2\kappa_1\log(\kappa_1\exp(\beta t)+1)\right.\\
&+\left.\phantom{\frac{}{}}\beta^2\kappa_1 (\kappa_1\exp(\beta t)-\beta t)+\kappa_1\kappa_2\alpha-\kappa_5\right)u_{(0)},
\end{aligned}
\end{equation}
and integrating the approximate invariant surface conditions \eqref{AISC}, we get the following representation of the solution:
\begin{equation}
\begin{aligned}
 u_{(0)}(t,x)&=\frac{U_0(x)}{\kappa_1\exp(\beta t)+1},\\
 u_{(1)}(t,x)&=\frac{1}{\kappa_1\exp(\beta t)+1}\left(-\frac{\exp\left(-\beta t-\frac{\alpha}{2}x\right)}{\beta}
\frac{\kappa_3\sin\left(\frac{\delta}{2} x\right)+\kappa_4\cos\left(\frac{\delta}{2} x\right)}{U_0(x)}\right.\\
 &-\frac{1}{{\beta\kappa_1(\kappa_1\exp(\beta t)+1)}}\left(\beta^2\kappa_1(\kappa_1\exp(\beta t)(2\log(\kappa_1\exp(\beta t)+1)\right.\\
 &-\left.\beta t)-1)-\kappa_1\kappa_2\alpha+\kappa_5\right)U_0(x)\\
 &+4\left.\frac{\kappa_2}{\beta}(\log(\kappa_1\exp(\beta t)+1)-\beta t)U_0^\prime(x)+U_1(x)\right).
\end{aligned}
\end{equation}
The associated approximate Q-conditional solution has the form:
\begin{equation}
\label{sol3a}
\begin{aligned}
 u(t,x;\varepsilon)&=\frac{8\beta}{(\alpha^2+\delta^2)(\kappa_1\exp(\beta t)+1)}\\
 &+\varepsilon\left(-\frac{\alpha^2+\delta^2}{8\beta^2}\exp\left(-\beta t-\frac{\alpha}{2}x\right)
\left(\kappa_3\sin\left(\frac{\delta}{2} x\right)+\kappa_4\cos\left(\frac{\delta}{2} x\right)\right)\right.\\
 &-\frac{8\exp(\beta t)}{(\alpha^2+\delta^2)(\kappa_1\exp(\beta t)+1)^2}(\beta^2\kappa_1(2\log(\kappa_1\exp(\beta t)+1)\\
 &-\left.\beta t+1)+\kappa_1\kappa_2\alpha-\kappa_5\right)\\
 &+\left.\frac{\exp\left(-\left(\alpha+\sqrt{\frac{\alpha^2-\delta^2}{2}}\right)\frac{x}{2}\right)}{\kappa_1\exp(\beta t)+1}\left(c_1\exp\left(\sqrt{\frac{\alpha^2-\delta^2}{2}}x\right)+c_2\right)\right).
\end{aligned}
\end{equation}
A plot of this solution is displayed in Figure~\ref{fig3a}.

Another case of admitted Q-conditional approximate symmetries has the approximate generators (and, consequently, the representation of the solution) with the same expression  as \eqref{gen112} and \eqref{rep112}. The approximate Q-conditional solution we find here reads
\begin{equation}
\begin{aligned}
 u(t,x;\varepsilon)&=\frac{8\beta}{\alpha^2+\delta^2}+\varepsilon\left(\frac{8\beta}{\alpha^2+\delta^2}\frac{\kappa_2}{\kappa_1}\exp(\beta t)+\exp\left(\left(\frac{\kappa_1(\alpha^2+\delta^2)}{8\beta}\right.\right.\right.\\
 &+\left.\left.\alpha+\sqrt{\frac{(\kappa_1(\alpha^2+\delta^2)+8\alpha\beta)^2}{64\beta^2}-\frac{\alpha^2+\delta^2}{2}}\right)\frac{\kappa_1 t-x}{2}\right)\times\\
 &\times\left. \left(c_1\exp\left(\sqrt{\frac{(\kappa_1(\alpha^2+\delta^2)+8\alpha\beta)^2}{64\beta^2}-\frac{\alpha^2+\delta^2}{2}}(-\kappa_1 t+x)\right)+c_2\right)\right).
\end{aligned}
\end{equation}

Finally, for the approximate generators
\begin{equation}
\begin{aligned}
 \xi_{(0)}&=0,\qquad \eta_{(0)}=-\beta u_{(0)},\\
 \xi_{(1)}&=4\exp(-\beta t)\left(\kappa_{1}\sin(\delta x)+\kappa_{2}\cos(\delta x)+\kappa_3\right)\\
 &+4\exp\left(\beta t+\frac{\alpha}{2}x\right)\left(\kappa_4\sin\left(\frac{\delta }{2}x\right)+\kappa_5\cos\left(\frac{\delta }{2}x\right)\right)u_{(0)}^2,\\
 \eta_{(1)}&=\exp\left(-3\beta t-\frac{\alpha}{2}x\right)\frac{\kappa_6\sin\left(\frac{\delta }{2}x\right)+\kappa_7\cos\left(\frac{\delta }{2}x\right)}{u_{(0)}}\\
 &-\exp(-\beta t)\left(\phantom{\frac{}{}}(\kappa_{1}\alpha+\kappa_2\delta)\sin(\delta x)-(\kappa_1\delta-\kappa_2\alpha)\cos(\delta x)\right.\\
 &+\left.\phantom{\frac{}{}}\kappa_3\alpha-\kappa_8\right)u_{(0)}-\beta^2u_{(0)}-\exp\left(\beta t+\frac{\alpha}{2} x\right)\times
\\
 &\times\left(\phantom{\frac{}{}}(\kappa_{4}\alpha+\kappa_5\delta)\sin(\delta x)-(\kappa_4\delta-\kappa_5\alpha)\cos(\delta x)\right)u_{(0)}^3,
\end{aligned}
\end{equation}
and 
\begin{equation}
\begin{aligned}
 \xi_{(0)}&=\beta\kappa_1\exp(-\beta t),\qquad \eta_{(0)}=-\beta u_{(0)},\\
 \xi_{(1)}&=(\kappa_2-\kappa_1\beta^3 t)\exp(-\beta t)+\kappa_1(\kappa_3-\kappa_1\alpha\beta^2)\exp(-2\beta t)\\
 &-\beta^2\frac{\alpha^2+\delta^2}{8}\kappa_1^3\exp(-3\beta t),\\
 \eta_{(1)}&=-\beta^3 \kappa_1^2\exp(-2\beta t)+\left(\frac{\beta^2(\alpha^2+\delta^2)}{4}\kappa_1^2\exp(-2\beta t)-\kappa_3\exp(-\beta t)-\beta^2\right)u_{(0)},
\end{aligned}
\end{equation}
after computing the corresponding representations of the solutions and obtaining the reduced ordinary differential equations, in
general we do not find real valued solution, so that we omit to write them. 

\section{Concluding remarks}
\label{sec4}
In this paper, by adopting a recently proposed approach to approximate Lie symmetries of differential equations involving \emph{small terms}, we introduced a consistent approach for investigating
approximate Q-conditional symmetries of partial differential equations. We applied the general framework to
a hyperbolic version of a reaction-diffusion-convection equation \cite{Murray}. A preliminary analysis of first order approximate Q-conditional symmetries of this equation have been recently discussed
\cite{GorgoneOliveri-EJDE}. Here, we classified a wide set of approximate Q-conditional invariant solutions, and plotted some of them by suitably choosing the values of the involved parameters. In the plots we observe that as $t$ increases the initial profile of the solution is damped and a
uniform solution reached. 

The approach here used is susceptible of being generalized, by considering
system of partial differential equations, investigating approximate Q-conditional symmetries of 
$p$th type \cite{Cherniha-JPA2010}, or developing approximate symmetry theories with a multiple scale approach.

As a last remark, we observe that approximate Lie symmetries of differential equations can be considered also in the framework of the inverse Lie problem, that is the problem of characterizing the form of differential equations by requiring the invariance with respect to a given Lie algebra of point symmetries. As far as the inverse Lie problem is concerned, in recent years, \emph{Lie remarkable equations} 
\cite{MOV-JMAA,MOV-TMP,MOSV,GO-JGP}, \emph{i.e.}, differential equations uniquely determined by the algebra of their point symmetries, have been defined and characterized. Therefore, a possible perspective could be that of considering suitable approximate Lie algebras and determine the differential equations containing small terms uniquely characterized by them at a fixed order of approximation.

\section*{Acknowledgments}
Work partly supported by GNFM of ``Istituto Nazionale di Alta Matematica'', and by local grants of the 
University of Messina.

\end{document}